\documentclass{emulateapj}

\newcommand{\sqcm}{${\rm cm^{-2}}$}
\newcommand{\cubcm}{${\rm cm^{-3}}$}
\newcommand{\mum}{${\rm \mu m}$}
\newcommand{\kms}{${\rm km~s^{-1}}$} 
\newcommand{\waven}{${\rm cm^{-1}}$}
\newcommand{\thirteenco}{${\rm ^{13}CO}$}
\newcommand{\twelveco}{${\rm ^{12}CO}$}

\newcommand{\eighteenco}{${\rm C^{18}O}$}
\newcommand{\bdop}{$b_{\rm D}$}

\newcommand{\irsnine}{NGC 7538 : IRS9}

\shorttitle{Detection of the Methane Stretch Mode}
\shortauthors{Boogert, Blake, \& \"{O}berg}

\received{2004 March 4, Accepted 2004 July 8; Scheduled for
publication in ApJ 615, 2004 November 1}

\begin{document}

\title{Methane Abundance Variations toward the Massive Protostar NGC
7538 : IRS9\footnote{Some of the data presented herein were obtained
at the W.M. Keck Observatory, which is operated as a scientific
partnership among the California Institute of Technology, the
University of California and the National Aeronautics and Space
Administration. The Observatory was made possible by the generous
financial support of the W.M. Keck Foundation.}
\footnote{Based on observations with ISO, an ESA project with
instruments funded by ESA Member States (especially the PI countries:
France, Germany, the Netherlands and the United Kingdom) and with the
participation of ISAS and NASA.}}

\author{A. C. A. Boogert\altaffilmark{3}, 
        G. A. Blake \altaffilmark{4,5},
        K. \"{O}berg\altaffilmark{5}}

\altaffiltext{3}{California Institute of Technology, Division of
              Physics, Mathematics and Astronomy 105-24, Pasadena, CA
              91125; acab@astro.caltech.edu}

\altaffiltext{4}{California Institute of Technology, Division of
              Geological and Planetary Sciences 150-21, Pasadena, CA
              91125}

\altaffiltext{5}{California Institute of Technology, Division of
              Chemistry and Chemical Engineering, Pasadena, CA
              91125}

\begin{abstract}
Absorption and emission lines originating from the $\nu _3$ C-H
stretching manifold of gas phase CH$_4$ were discovered in the high
resolution ($R$=25,000) infrared L band spectrum along the line of
sight toward \irsnine.  These observations provide a diagnostic of the
complex dynamics and chemistry in a massive star forming region. The
line shapes resemble P Cygni profiles with the absorption and emission
components shifted by $\sim 7$ \kms\ with respect to the systemic
velocity. Similar velocity components were observed in CO at 4.7 \mum,
but in contrast to CH$_4$, the CO shows deep absorption due to a high
velocity outflow as well as absorption at the systemic velocity due to
the cold outer envelope. It is concluded that the gas phase CH$_4$
abundance varies by an order of magnitude in this line of sight: it is
low in the envelope and the outflow (X[CH$_4$]$<0.4\times 10^{-6}$),
and at least a factor of 10 larger in the central core.  The discovery
of solid CH$_4$ in independent ground and space based data sets shows
that methane is nearly entirely frozen onto grains in the envelope.
It thus appears that CH$_4$ is formed by grain surface reactions,
evaporates into the gas phase in the warm inner regions of
protostellar cores and is efficiently destroyed in shocks related to
outflows.
\end{abstract}


\section{Introduction}~\label{sec:intro}

Among the suite of molecules observed in interstellar and
circum-protostellar media\footnote{see the compilation at\\
\url{http://www.cv.nrao.edu/$\sim$awootten/allmols.html}}, methane is
relatively poorly studied.  CH$_4$ has no permanent dipole moment, and
therefore cannot be observed by pure rotational transitions at radio
wavelengths.  Its strongest fundamental ($v=1-0$) ro-vibrational
transitions, i.e.  the $\nu_3$ C-H stretching and the $\nu_4$ C-H
bending modes, occur at 3.32 and 7.67 \mum, respectively. Strong
telluric absorption at these wavelengths hinders interstellar CH$_4$
studies.  Gas and solid phase CH$_4$ have been positively identified
at 7.67 \mum\ in three sight-lines: the massive protostars NGC 7388 :
IRS9, W~33A, and GL 7009 S \citep{lac91, boo96, boo97, boo98, dar98}.
CH$_4$ was detected only in the solid state in low resolution {\it
Infrared Space Observatory (ISO)} observations at 7.67 \mum\ toward
$\sim 30$ low mass protostars \citep{whi00, gue02, ale03}, the
Galactic Center \citep{chi00}, and an external galactic nucleus
\citep{spo01}. A recent low resolution {\it Spitzer Space Telescope}
spectrum shows absorption by solid CH$_4$ toward the young, low mass
protostar IRAS 08242+5100 \citep{nor04, boo04}.

Observations of CH$_4$ provide insight into the basic principles of
astrochemical networks.  In the presence of atomic C and H, CH$_4$ is
rapidly formed on cold grains, just as H$_2$O is formed from atomic O
and H (e.g.  \citealt{bro88}).  Perhaps surprisingly, the observed gas
and solid state CH$_4$ abundances are low, not more than a few percent
with respect to CO and H$_2$O \citep{lac91, boo98}.  This points to
relatively low atomic C abundances at the time of CH$_4$ formation,
when most C is already locked up in CO. In this same scenario the high
CH$_3$OH abundances in several lines of sight may be explained by
hydrogenation of abundantly present CO \citep{dar99, pon03}, and large
CO$_2$ abundances by oxidation of CO. The dominance of a CO- and
O-based, rather than C-based, chemistry is stressed by the profiles of
solid CO$_2$ absorption bands which show an intimate mixture of
CO$_2$, CH$_3$OH, and H$_2$O in the ices \citep{ger99, boo00}.
Nevertheless, even at relatively low abundances methane is proposed to
be the starting point of a rich chemistry leading to the complex
organic molecules observed in dense clouds like the TMC-1 ridge
\citep{mar00}. To further test the role of CH$_4$ in interstellar
chemistry, basic diagnostics, such as gas phase abundances, gas/solid
phase abundance ratios, gas phase temperatures and abundance gradients
need to be measured in more lines of sight. 

In this work we report the discovery of absorption and emission
features arising from the 3.32 \mum\ C-H stretching mode of gas and
solid phase CH$_4$.  A previous attempt to detect CH$_4$ at this
wavelength was unsuccessful and indicated an abundance of $< 10^{-6}$
with respect to H in the Orion BN object \citep{kna85}. We study the
line of sight toward the massive protostar NGC 7538 : IRS9, because
its high radial velocity of $V_{\rm lsr}=-57$ \kms\
(e.g. \citealt{tak00}) facilitates the separation of telluric and
interstellar lines.  NGC 7538 : IRS9 is a well studied massive
protostar, known for its large ice and gas columns \citep{whi96,
mit90} and a prominent molecular outflow (Mitchell, Maillard, \&
Hasegawa 1991).  The R(0) and R(2) lines of the bending mode of gas
phase CH$_4$ and solid state CH$_4$ were detected in this source by
\citet{lac91}. Unambiguous confirmations of these detections were
obtained with space based observations \citep{boo96, boo98}. Here we
independently confirm the presence of CH$_4$ by detecting many lines
from the 3.32 \mum\ stretching mode as well as the ice band. The high
spectral resolution of these observations permits a line profile
analysis in which we determine gas phase conditions in the various
environments along the line sight, such as the envelope, the hot core,
and the outflow. This, in turn, elucidates the formation and
destruction pathways of CH$_4$ in these different environments.

The Keck/NIRSPEC and ISO/SWS observations are discussed in
\S~\ref{sec:obs}, and presented in \S~\ref{sec:feat}.  A profile
analysis of the gas phase absorption and emission CH$_4$ lines, in
conjunction with previously reported high resolution CO observations,
is presented in \S\S~\ref{sec:ch4} and ~\ref{sec:co}. The ice band
profile is discussed and compared with the C-H bending mode in
\S~\ref{sec:ice}.  The implications of these results on the formation
and evolution history of the CH$_4$ molecule in protostellar
environments are discussed in \S~\ref{sec:disc1}. The origin of the
low velocity, warm gas, responsible for the P Cygni-like line
profiles, is further constrained in \S~\ref{sec:disc2}.  We conclude
with future prospects in \S~\ref{sec:concl}.

\begin{figure*}
\includegraphics[angle=90, scale=0.70]{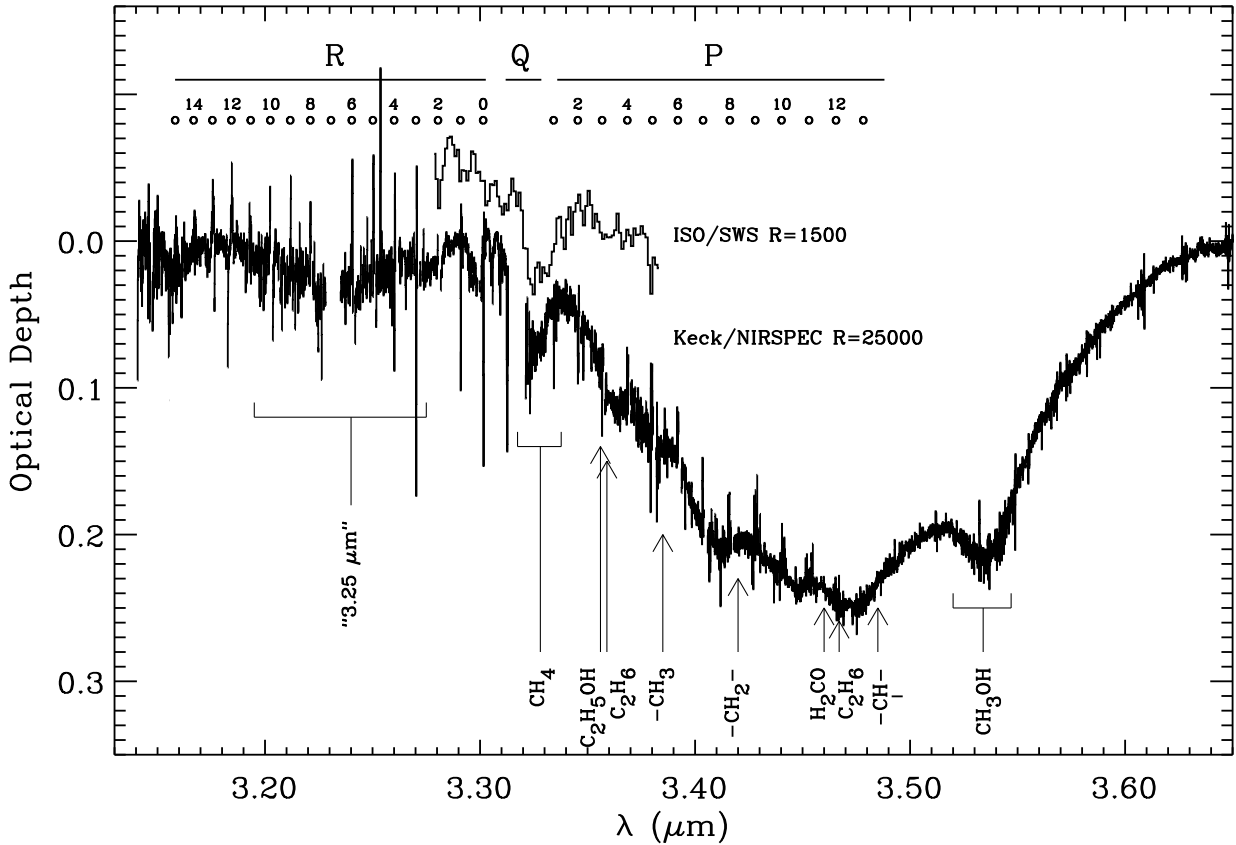}
\caption{The R=25,000 Keck/NIRSPEC L band spectrum of \irsnine\ on an
optical depth scale, after correction for H$_2$O ice absorption. The
broad, deep absorption is the well known, yet unidentified, ``3.47
\mum'' absorption band. The presence of absorption by the C-H
stretching mode of solid CH$_4$ is confirmed in the R=1,500 ISO/SWS
spectrum.  P, Q, and R-branch wavelengths of gas phase CH$_4$ are
indicated at the top, where the numbers refer to the lower rotational
$J$ values. The gas phase lines are further highlighted in
Fig.~\ref{f:gas}.  Additional solid state absorption features are
superposed on the 3.47 \mum\ band. The most secure identification is
that of methanol \citep{all92}. Other possible identifications are
indicated by arrows, but these are not further discussed in this
paper: ethanol \citep{all92}, ethane \citep{bou98}, formaldehyde
\citep{sch96}, and diffuse medium aliphatic C-H stretching modes
\citep{san91}. The weak, broad inflection near 3.25 \mum\ is possibly
due to absorption by aromatic hydrocarbons
\citep{bro99}.}~\label{f:obs}
\end{figure*}

\vspace{20pt}

\section{Observations}~\label{sec:obs}

\subsection{Keck/NIRSPEC R=25,000 Spectroscopy}~\label{sec:obsns}

The high resolution L band spectrum of the massive protostar \irsnine\
was obtained with the NIRSPEC spectrometer \citep{mcl98} at the Keck
II telescope atop Mauna Kea. Observations were made on five different
nights with clear and dry weather conditions: UT 2002 December 17 and
18, 2003 July 10 and 11, and 2003 November 2.  NIRSPEC was used in the
echelle mode with the 0.43$\times 24''$ slit, providing a resolving
power of $R=\lambda/\Delta \lambda=25,000$ ($\sim$12 \kms). The
2.9--4.1 \mum\ region was almost fully covered in five echelle
settings, each covering five Nyquist sampled wavelength orders of
$\sim 0.05$ \mum\ on the 1024$^2$ pixel array. Two small regions
centered on 2.97 and 3.12 \mum\ were not covered.  The on-source
integration time per setting was 20 minutes. Several CH$_4$
ro-vibrational lines as well as the solid CH$_4$ band at 3.32 \mum\
were observed multiple times to verify their reality.  At $V_{\rm
lsr}=-57$ \kms\ for \irsnine, the Doppler velocity with respect to
earth was $-$53.5, $-$82.4, and $-$64.0 \kms\ for the three observing
runs. This proved to be sufficient to separate interstellar and
telluric CH$_4$ lines and guarantees their reality. The spectra
presented in this paper are an average of all the data.  The data were
reduced in a standard way, using IDL routines (Boogert, Blake, \&
Tielens 2002). Atmospheric absorption features were divided out using
the standard star HR 8585 (A1V), which is bright ($V$=3.78) and
reasonably close to \irsnine\ (an airmass of 1.20-1.30 versus 1.34 for
\irsnine). The spectral shape and hydrogen absorption features in the
standard star were divided out with a Kurucz model atmosphere. The
hydrogen absorption profiles in the Kurucz model are accurate to
1-2\%, and residuals can be seen near strong lines. This does not
affect the region of main interest for this paper (3.15--3.65 \mum),
except perhaps near 3.297 \mum\ due to the HI Pf $\delta$
line. Overall, a good telluric correction was achieved, resulting in
signal-to-noise values of $\sim$110 in the region of CH$_4$ at 3.32
\mum, $\sim 40$ at the bottom of the H$_2$O ice band at 3.0 \mum\ and
$\sim 200$ at wavelengths $>3.5$ \mum.  Strong atmospheric lines, i.e.
with less than 50\% of the maximum transmission in each setting, leave
residuals and were removed from the final spectrum.  The spectra were
wavelength calibrated on the atmospheric emission lines, using an
atmospheric model spectrum (P. van der Valk and P. Roelfsema,
priv. comm.). All orders were subsequently combined by applying
relative multiplication factors.  For the analysis of the emission
lines (\S\ref{sec:ch4}), the spectrum is flux calibrated by scaling to
the continuum brightness of the data presented in \citet{whi96}.
Finally, in conjunction with the CH$_4$ features, we analyze
Keck/NIRSPEC M band CO lines in \irsnine, previously published in
\citet{boo02}.

\vspace{20pt}

\subsection{ISO/SWS R=1,500 Spectroscopy}~\label{sec:obssws}

The 3.26-3.38 \mum\ spectral region of \irsnine\ was observed multiple
times in the high resolution R=1,500 AOT06 mode of the Short
Wavelength Spectrometer on board the Infrared Space Observatory
(ISO/SWS; \citealt{kes96}; \citealt{gra96}). From the ISO archive we
have selected the longest integrations in this wavelength range,
corresponding to TDT numbers 39002336 (UT 11 Dec 1996), 75101049 (UT
05 Dec 1997), and 85200455 (UT 16 Mar 1998). All observations were
part of the ICE\_BAND/DWHITTET guaranteed time program, and have not
been previously published. The combined on-target integration time in
the 3.26-3.38 \mum\ region amounts to 5219 seconds. The individual
data sets were processed with SWS Interactive Analysis and calibration
files version 10.1 in April 2003 at SRON/Groningen.  The three AAR
products were combined and the detector scans were aligned to the mean
using second order polynomials.  Then, in this highly oversampled
spectrum, data points deviating more than 3$\sigma$ from the mean per
resolution element were removed. Finally, the data points were
averaged and re-sampled to an R=1,500 Nyquist sampled grid.  The
resulting S/N is 100, but significantly poorer at the outer 0.015
\mum\ edges.

\vspace{20pt}

\section{Results}~\label{sec:res}

\subsection{Features in the 3.1-3.6 \mum\ Spectral Region}~\label{sec:feat}

\begin{figure*}
\includegraphics[angle=90, scale=0.75]{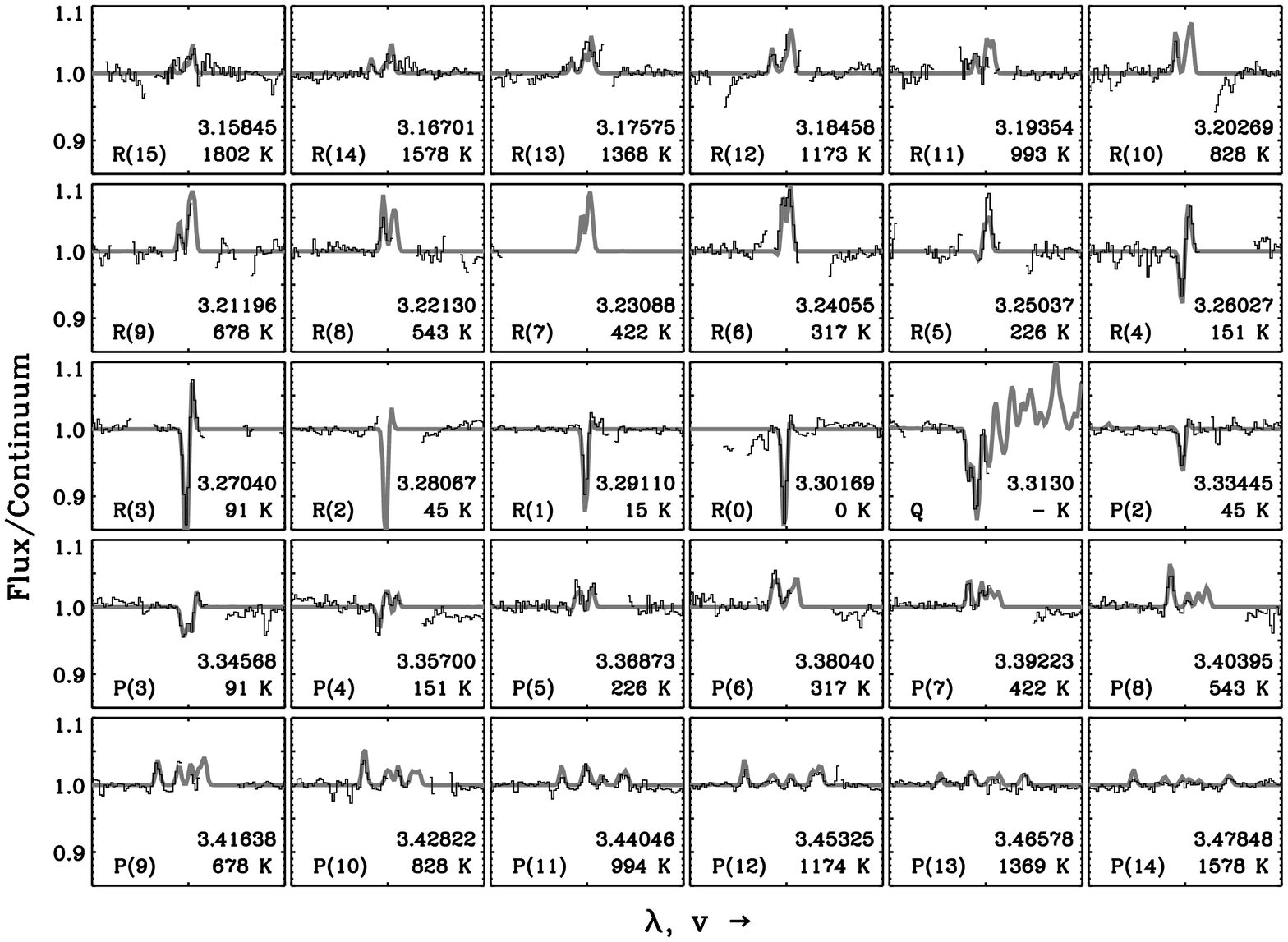}
\caption{Magnification of the Keck/NIRSPEC spectrum of \irsnine\ in
spectral regions surrounding gas phase CH$_4$ ro-vibrational lines
($\Delta \lambda$=0.006 \mum\ or $\Delta v$=530 \kms).  This is the
same data as shown in Fig.~\ref{f:obs}, however on a flux/continuum
scale.  The gaps in the spectra are regions with less than 50\%
atmospheric transmission that were removed from the data. The thick
gray line superposed on each observed spectrum depicts a two component
LTE model, composed of a 6 \kms\ blue-shifted absorption component
($T=55$ K, $N=2.5\times 10^{16}$ \sqcm), and an 8 \kms\ red-shifted
emission component ($T=600$ K, $N=5.8\times 10^{16}$ \sqcm;
Table~\ref{t:col}). The ro-vibrational transition, the lower
rotational energy level (in K), and the center wavelength (in \mum)
are indicated in the lower left and right corners of each panel
respectively. Note that CH$_4$ emission is detected up to the highest
excitation levels, i.e. up to R(15) and P(14). }~\label{f:gas}
\end{figure*}

The L band spectrum of \irsnine\ is dominated by the deep absorption
band of the O-H stretching mode of H$_2$O ice \citep{ wil82,
whi96}. Its peak optical depth at 3.0 \mum\ is $\sim 4.5$. The most
prominent substructure in the long wavelength wing of H$_2$O is the
well known, yet unidentified, 3.47 \mum\ absorption band \citep{all92,
bro96, dar02}. We applied a smooth spline continuum to extract this
band on an optical depth scale (Fig.~\ref{f:obs}). The results were
verified by applying low order polynomials to continuum regions
described in previous studies (e.g. Brooke, Sellgren, \& Geballe
1999). The absorption shape is in good agreement with the low
resolution spectrum of \citet{all92}, as is the presence of a sub-band
at 3.53 \mum, attributed to the C-H stretching mode of solid CH$_3$OH.
A previously undetected absorption band is apparent by a downturn in
the continuum slope at 3.32 \mum.  Although it is in a spectral region
badly affected by the telluric CH$_4$ Q-branch, its presence is
confirmed in Keck/NIRSPEC observations on different dates. In
addition, a feature at the same wavelength is found in the space based
ISO/SWS spectrum (Fig.~\ref{f:obs}). We thus conclude that this
feature is real and is due to the stretching mode of solid CH$_4$ (\S
\ref{sec:ice}). Not immediately obvious in Fig.~\ref{f:obs} but
clearly detected upon close inspection (Fig.~\ref{f:gas}) are P, Q,
and R-branch lines of gas phase CH$_4$.  It is the first time that the
3.32 \mum\ ice and gas features are detected in a circum-protostellar
environment.

A number of additional features, not further discussed in this paper,
are visible in the high resolution Keck/NIRSPEC spectrum. Although
they are reproduced in observations at different dates, they are
entirely new and need to be confirmed in a larger sample of sources,
or, as for solid CH$_4$, in independent data.  A weak feature at 3.36
\mum\ coincides with the C-H stretch modes of ethanol \citep{all92}
and ethane (Boudin, Schutte, \& Greenberg 1998).  Interestingly,
another band of ethane at 3.47 \mum\ seen in laboratory ices appears
to be present in \irsnine\ as well.  Finally, weak absorptions at 3.47
and 3.41 \mum\ are also close to the C-H stretch mode of single bonded
carbon in CH and CH$_2$ groups in aliphatic hydrocarbons, normally
seen in diffuse lines of sight only (e.g. \citealt{pen94}).

\subsection{Gas Phase Line Profiles at R=25,000}~\label{sec:gas}

\subsubsection{Gas Phase CH$_4$}~\label{sec:ch4}

Ro-vibrational absorption and emission lines of gaseous CH$_4$ are
clearly detected toward \irsnine\ (Fig.~\ref{f:gas}). Although the
telluric CH$_4$ lines are deep, the large blue-shift of \irsnine\ with
respect to the earth (\S~\ref{sec:obsns}) separates many interstellar
lines well.  The lines appear mostly in absorption up to lower energy
levels with $J=3$, and entirely in emission above $J=6$. Intermediate
$J$ levels have P Cygni-like line profiles, i.e. red-shifted emission
and blue-shifted absorption.  Such line profiles are usually
associated with expanding envelopes (see \S\ref{sec:disc2}, however).
Note that for $J>4$ multiple emission and absorption peaks appear for
each $J$ level. These line splittings are intrinsic to the CH$_4$
molecule and result from rotation-vibration interactions with the spin
manifolds of the four equivalent H nuclei.  Thus E--, F-, and A--type
multiplets occur with spin statistical weights of $g_{\rm s}$=2, 3,
and 5 respectively.

\begin{deluxetable}{lcccc}
\tablecolumns{5}
\tablewidth{0pc}
\tablecaption{CH$_4$ Emission Line Strengths toward \irsnine}~\label{t:ch4em}
\tablehead{
\colhead{$\Delta v(J)$}& \colhead{$\lambda$$^{\rm a}$} & \colhead{lines$^{\rm b}$} & 
         \colhead{$F'_{\rm J}/F_{\rm J}$$^{\rm c}$} & \colhead{$F'_{\rm J}$$^{\rm d}$}\\
\colhead{} & \colhead{\mum} & \colhead{}   &   \colhead{\%} & \colhead{10$^{-21}$ ${\rm W/cm^2}$}}\\
\startdata
R(15) & 3.15812--3.15864 &  11   & 84    & 5.2 (1.1) \\
      & 3.15767--3.15792 &   3   & 13    & $<$3.3    \\
R(14) & 3.16677--3.16720 &  13   & 73    & 3.8 (0.7) \\
      & 3.16626--3.16665 &   5   & 26    & 1.5 (0.4) \\
R(13) & 3.17545--3.17595 &  13   & 81    & 9.0 (0.8) \\
      & 3.17498--3.17534 &   4   & 16    & 4.3 (0.8) \\
R(12) & 3.18435--3.18485 &  14   & 74    & 13.6 (3.0)\\
      & 3.18390--3.18428 &   4   & 26    & 4.8 (0.5) \\
R(11) & 3.19295--3.19325 &   2   & 21    & 3.8 (1.5) \\
R(10) & 3.20211--3.20253 &   3   & 34    & 3.5 (1.0) \\
R(9)  & 3.21178--3.21229 &   6   & 78    & 10.7 (3.0)\\
      & 3.21146--3.21179 &   2   & 22    & 3.0 (0.8) \\
R(8)  & 3.22087--3.22118 &   3   & 47    & 4.8 (0.8) \\
R(6)  & 3.24014--3.24080 &   6   & 100   & 25.1 (1.5)\\
R(5)  & 3.25019--3.25064 &   4   & 100   & 12.5 (1.4)\\
R(4)  & 3.26017--3.26054 &   4   & 100   & 7.5 (1.0) \\
R(3)  & 3.27035--3.27063 &   3   & 100   & 6.6 (0.5) \\
R(1)  & 3.29105--3.29127 &   1   & 100   & 1.5 (0.5) \\
R(0)  & 3.30165--3.30187 &   1   & 100   & 1.5 (0.6) \\
P(2)  & 3.33435--3.33463 &   2   & 100   & 1.8 (0.4) \\
P(3)  & 3.34573--3.34600 &   3   & 45    & 2.1 (0.5) \\
P(4)  & 3.35708--3.35736 &   1   & 23    & 1.9 (0.6) \\
      & 3.35679--3.35703 &   3   & 77    & 1.9 (0.9) \\
P(5)  & 3.36860--3.36901 &   2   & 45    & $\geq$4.8 \\
      & 3.36814--3.36860 &   2   & 55    & 4.5 (0.6) \\
P(6)  & 3.37967--3.38022 &   3   & 48    & 7.8 (0.8) \\
P(7)  & 3.39182--3.39230 &   2   & 42    & $>5$      \\
      & 3.39138--3.39176 &   2   & 32    & 3.5 (0.4) \\
P(8)  & 3.40379--3.40411 &   1   & 14    & $\geq$1.6 \\
      & 3.40318--3.40370 &   3   & 48    & 6.8 (0.4) \\
P(9)  & 3.41577--3.41613 &   2   & 18    & $>2.7$    \\
      & 3.41509--3.41560 &   2   & 22    & 4.1 (0.4) \\
P(10) & 3.42719--3.42772 &   3   & 35    & 4.2 (0.7) \\
P(11) & 3.44128--3.44172 &   3   & 30    & 3.2 (0.4) \\
      & 3.44072--3.44109 &   1   & 11    & 1.6 (0.2) \\
      & 3.44009--3.44057 &   3   & 37    & 3.9 (0.2) \\
      & 3.43944--3.43982 &   2   & 22    & 2.0 (0.2) \\
P(12) & 3.45380--3.45441 &   4   & 43    & 4.7 (0.3) \\
      & 3.45313--3.45347 &   2   & 13    & 1.4 (0.3) \\
      & 3.45252--3.45296 &   2   & 16    & 1.1 (0.4) \\
      & 3.45175--3.45208 &   3   & 27    & 1.9 (0.3) \\
P(13) & 3.46664--3.46711 &   3   & 23    & 1.1 (0.4) \\
      & 3.46566--3.46629 &   3   & 31    & 0.8 (0.3) \\
      & 3.46501--3.46548 &   3   & 28    & 1.8 (0.3) \\
      & 3.46406--3.46457 &   2   & 17    & 0.8 (0.3) \\
P(14) & 3.47952--3.48004 &   3   & 22    & 1.9 (0.4) \\
      & 3.47814--3.47899 &   4   & 35    & 0.6 (0.3) \\
      & 3.47757--3.47796 &   2   & 16    & $<0.8$    \\
      & 3.47659--3.47707 &   3   & 27    & 1.7 (0.4) \\
\noalign{\smallskip} 
\tableline
\multicolumn{5}{p{7cm}}{ $^{\rm a}$ integration region
after correction for 8 \kms\ red-shift}\\
\multicolumn{5}{p{7cm}}{ $^{\rm b}$ number of blended lines in
integration interval}\\
\multicolumn{5}{p{7cm}}{ $^{\rm c}$ percentage of emission with
respect to total emission for this upper $J$ level}\\
\multicolumn{5}{p{7cm}}{ $^{\rm d}$ $1\sigma$ errors in brackets}\\
\enddata
\end{deluxetable}

The observed lines are analyzed using the 2001 update of the HITRAN
2000 database \citep{rot03, bro03}.  All HITRAN transitions
contributing to each observed interstellar line are identified and the
number of transitions responsible for at least 90\% of the line
strength is indicated in column 3 of Tables 1 and~\ref{t:ch4ab}. For
example, at the resolution of our observations, the P(11) transition
is split in 4 separate emission lines, which are composed of some 9
strong transitions. The contribution of each observed line to the
total emission or absorption per $J$ level is indicated in column 4 of
Tables 1 and~\ref{t:ch4ab} and is calculated in the optically thin
limit as

\begin{equation}
 \frac{F'_J}{F_J}=\frac{\sum_{i}g_{iJ}A_{iJ}}{\sum_{n}g_{nJ}A_{nJ}}
 \times 100\%{\rm ,}
\end{equation}

\noindent where $g_{iJ}$ is the product of the rotation (2$J$+1) and
nuclear spin ($g_{\rm s}$) statistical weights and $A_{iJ}$ is the
corresponding Einstein coefficient for transition $i$ at rotational
level $J$ .  The summation over $i$ is for all transitions in the
wavelength range of an observed emission line (column 2 in Tables 1
and~\ref{t:ch4ab}) while the summation over $n$ is for all transitions
at rotational level $J$. For several $J$ levels only part of the
emission or absorption is observed as some components lie in poor
atmospheric regions. For example, only half the R(8) line flux is seen
and for the Q-branch the clearly detected absorption component
represents only a few percent of the total Q-branch
absorption. Finally, the observed emission line fluxes $F'_J$ listed
in Table 1 are corrected for the underlying H$_2$O absorption using a
wide coverage ISO/SWS spectrum \citep{whi96}. The resulting continuum
flux of 1.5 Jy is corrected for foreground extinction using the
Galactic extinction law \citep{dra03} and the $A_{\rm V}$ derived in
\S\ref{sec:disc2}. Thus, the observed line fluxes are scaled to an
unextincted continuum flux of 23.7 Jy.

\begin{figure}
\includegraphics[angle=90, scale=0.49]{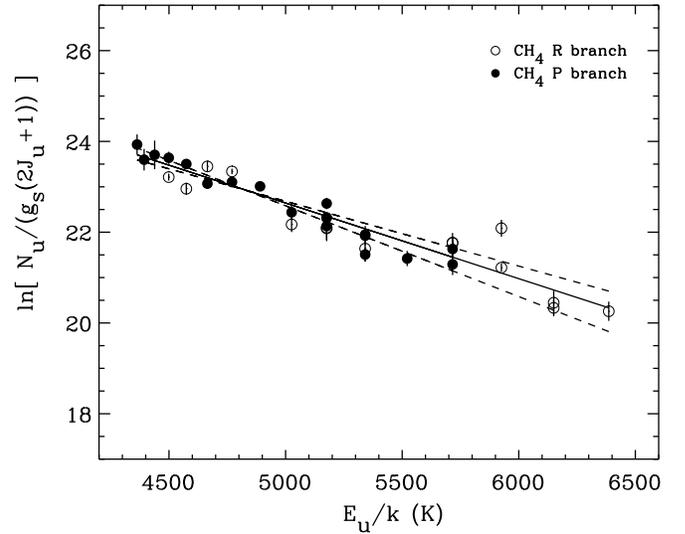}
\caption{Rotation diagram of CH$_4$ emission lines toward
\irsnine. P-branch lines are represented by filled circles and
R-branch lines by open circles. The solid line represents a
least-squares best fit to the data corresponding to a rotational
temperature of 600 K at a column density of 5.8$\times 10^{16}$ \sqcm\
in a solid angle of 4.3$\times 10^{-14}$ sterad. The dashed lines
represent extreme (3$\sigma$) physical parameters that still fit the
data: $T_{\rm rot}=$500 K and $N$(CH$_4$)=20$\times 10^{16}$ \sqcm,
and $T_{\rm rot}=$700 K and $N$(CH$_4$)=2.6$\times 10^{16}$
\sqcm.}~\label{f:rotch4}
\end{figure}

In order to derive basic physical parameters, we construct a rotation
diagram for the CH$_4$ emission lines. The Boltzmann equation of a
single transition $i$ at upper rotational level $J$ is 

\begin{equation}
\frac{N_{iJ}}{g_{iJ}}=\frac{N_{\rm tot}}{Q(T_{\rm
rot})}e^{-E_{iJ}/{\rm k}T_{\rm rot}}{\rm ,}~\label{e:bol}
\end{equation}

\noindent where $N_{iJ}$ and $E_{iJ}$ are the column density and the
energy of the upper level of a single multiplet component
respectively.  Assuming that collisional excitation is applicable,
$E_{iJ}$ includes a vibrational $v=1$ excitation energy of 4348 K
\citep{lid94}.  $N_{\rm tot}$ and $T_{\rm rot}$ are the total CH$_4$
column density and rotational temperature.  Throughout this paper we
take the ro-vibrational partition function $Q(T_{\rm rot})$ from the
HITRAN database, which is applicable at $T_{\rm rot}=70-3000$ K
\citep{fis03}. At lower temperatures we calculate $Q(T_{\rm rot})$
from lines in the HITRAN database. In the optically thin limit, the
flux $F'_{J}$ of a partial $J$ transition is obtained by summation
over all multiplet lines

\begin{equation}
F'_{J}=\sum_{i}F_{iJ}~\label{e:f}
\end{equation}

\noindent and thus the observed $F'_{J}$ relates to eq.~\ref{e:bol} as

\begin{equation}
\frac{4\pi F'_{J}}{{\rm hc}\nu_{J} \times \Omega \times
\sum_{i}g_{iJ}A_{iJ}}=\frac{N_{\rm tot}}{Q(T_{\rm
rot})}e^{-E_{J}/{\rm k}T_{\rm rot}}.~\label{e:rot}
\end{equation}

\begin{figure*}
\includegraphics[angle=90, scale=0.70]{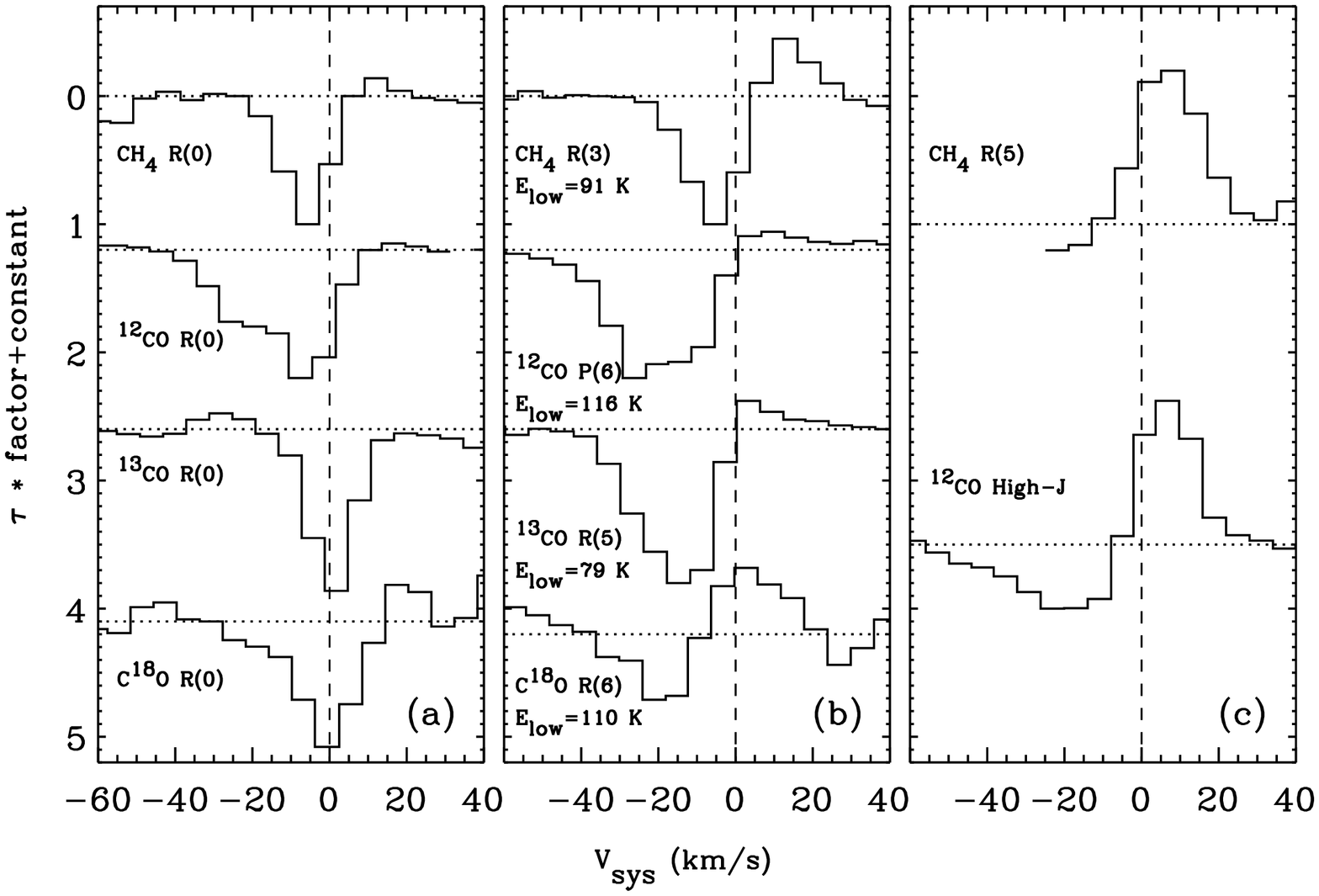}
\caption{Line profiles of CH$_4$ and CO isotopologues on a velocity
scale corrected for the earth and heliocentric velocity. Lines of
similar lower energy level are plotted: 0 K in panel (a), and $\sim
$100 K in panel (b).  For reference we have indicated the systemic
velocity with a vertical dashed line.  Note the very different profile
of CH$_4$ compared to CO lines. The --16 \kms\ outflow is absent in
CH$_4$, and the R(0) CH$_4$ line does not peak at the systemic
velocity, indicating that CO and CH$_4$ have different a chemical
origin and evolution (\S\ref{sec:disc1}). Panel (c) shows
emission-dominated CH$_4$ and CO line profiles. The high-$J$ CO line
profile toward \irsnine\ is an average over the individual lines
P(27), P(28), P(30), and P(34). The emission lines clearly peak at +8
\kms\ opposite to what is expected for P Cygni line profiles
(\S\ref{sec:disc2}).}~\label{f:velo}
\end{figure*}

\begin{table}
\caption{CH$_4$ Absorption Line Strengths toward
         \irsnine}~\label{t:ch4ab}
\center
\begin{tabular}{lccccc}
\noalign{\smallskip} 
\noalign{\smallskip} 
\tableline
\tableline
\noalign{\smallskip} 
$\Delta v(J)$& $\lambda$$^{\rm a}$ & lines$^{\rm b}$ & $F'_{\rm J}/F_{\rm J}$$^{\rm c}$ & eq. width$^{\rm d}$ \\
             & \mum &    &   \%  & $10^{-3}$ \waven    \\
\noalign{\smallskip}  
\tableline
\noalign{\smallskip} 
R(4)  & 3.26003--3.26031 &   3   & 77    & 10.0 (1.4) \\
R(3)  & 3.27019--3.27051 &   3   & 100   & 12 (1)     \\
R(1)  & 3.29092--3.29121 &   1   & 100   & 12.7 (0.4) \\
R(0)  & 3.30151--3.30179 &   1   & 100   & 17.6 (1.0) \\
Q     & 3.31234--3.31296 &   4   & 2     & 35 (6)     \\
P(2)  & 3.33423--3.33453 &   2   & 100   & 6.5 (0.8)  \\
P(3)  & 3.34530--3.34584 &   3   & 100   & 12.3 (0.7) \\
P(4)  & 3.35660--3.35694 &   3   & 61    & 5.9 (0.8)  \\
\noalign{\smallskip} 
\tableline
\multicolumn{5}{p{7cm}}{ $^{\rm a}$ integration region
after correction for 6 \kms\ blue-shift}\\
\multicolumn{5}{p{7cm}}{ $^{\rm b}$ number of blended lines in
integration interval}\\
\multicolumn{5}{p{7cm}}{ $^{\rm c}$ percentage of absorption with
respect to total absorption for this lower $J$ level}\\
\multicolumn{5}{p{7cm}}{ $^{\rm d}$ $1\sigma$ errors in brackets}\\
\end{tabular}
\end{table}

\noindent Here we make use of the fact that the upper energy level
$E_{J}$ and transition frequency $\nu_{J}$ are nearly the same for all
multiplet lines (to within 0.1\%). $\Omega$ is the size of the
emitting region, for which we assume the value of 4.3$\times 10^{-14}$
sterad derived from CO emission lines, two orders of magnitude smaller
than the solid angle corresponding to the 0.42$''$ wide slit
(\S~\ref{sec:disc2}). The rotation diagram constructed from the
observed CH$_4$ emission lines indicates a rotational temperature of
600$\pm 100$ K (Fig.~\ref{f:rotch4}). Although $T_{\rm rot}$ is well
constrained, the derived column density of $5.8^{+16}_{-3}\times
10^{16}$\sqcm\ is accurate to within a factor of $\sim$3 only
(3$\sigma$ errors).

To further illustrate the goodness of fit, to investigate possible
optical depth effects and to model the complex blend of CH$_4$
absorption and emission lines, we have also constructed single
temperature LTE models.  The level populations are calculated
according to eq.~\ref{e:bol}. Emission line fluxes and absorption
equivalent widths at any optical depth are calculated following
\citet{gen92}, and \citet{spi78} respectively, applying the Voigt
function for the line profiles. The observed line profiles are fitted
by adding a single temperature red-shifted emission model to a single
temperature, blue-shifted, absorption model. We find that a blue-shift
of 6 \kms\ and a red-shift of 8 \kms\ are needed, and that both
components are unresolved at the spectral resolution of 12 \kms.  The
validity of the temperature and column density derived from the
emission line rotation diagram is confirmed by the relatively low
model line optical depth; for example, at a Doppler parameter of
\bdop=1 \kms, $N_{\rm tot}=5.8\times 10^{16}$ \sqcm, and $T$=600 K,
the peak optical depth is 0.33 (in the R(8) line).  The absorption
column is well constrained to $N$(CH$_4$)=3$\pm 1\times 10^{16}$
\sqcm\ at $T_{\rm ex}$=55$\pm$15 K for intrinsic line widths of
\bdop=0.5--3 \kms\ (Table~\ref{t:col}). The narrowest lines are
optically thick ($\tau$=5 at \bdop=0.5 \kms), while they are optically
thin for \bdop$> 1 $ \kms.  We take \bdop=3 \kms\ as a preferred value
throughout this paper, because the larger line widths are usually
preferred toward \irsnine\ \citep{mit90, boo03}, but values of
\bdop$>3$ \kms\ are inconsistent with the unresolved CH$_4$ lines
observed.  The physical parameters derived from the CH$_4$ stretching
mode are in good agreement with the parameters from the bending mode
at 7.67 \mum\ in lower resolution spectra \citep{lac91, boo98}.

Finally, absorption or emission lines of the $^{13}$CH$_4$ and CH$_3$D
isotopes have not been detected. Study of the $^{13}$CH$_4$ lines is
complicated, because the isotope wavelength shift is comparable to the
separation of $J$ levels and thus the $^{13}$CH$_4$ and $^{12}$CH$_4$
lines are blended.  The column density upper limits are a factor of
3--7 smaller compared to the CH$_4$ column density. At
$^{12}$C/$^{13}$C and H/D isotope ratios of at least 50 in dense
clouds \citep{wil94, cec01}, these upper limits are not significant.

\subsubsection{Gas Phase CO}~\label{sec:co}

\begin{table}[b]
\caption{Gas Phase CO Absorption toward \irsnine}~\label{t:co}
\center
\begin{tabular}{lcc}
\noalign{\smallskip} 
\noalign{\smallskip} 
\tableline
\tableline
\noalign{\smallskip} 
$\Delta v(J)^{\rm a}$ & \multicolumn{2}{c}{eq. width$^{\rm b}$} \\
                      & \multicolumn{2}{c}{$10^{-3}$ \waven}    \\
\noalign{\smallskip} 
\cline{2-3}
\noalign{\smallskip} 
                      & \thirteenco & \eighteenco              \\
\noalign{\smallskip}  
\tableline
\noalign{\smallskip} 
R(18) & 2 (1)    & \ldots      \\
R(17) & 5 (2)    & \ldots      \\
R(16) & 3 (1.1)  & \ldots      \\
R(15) & 6.5 (2)  & \ldots      \\
R(13) & 10 (2)   & \ldots      \\
R(12) & 16 (3)   & \ldots      \\
R(11) & 12 (4)   & $<$4.5   \\
R(10) & 16 (3)   & $<$4.5   \\
R(9)  & 18 (5)   & 5 (1)    \\
R(7)  & 30 (8)   & 2.2 (1.0)\\
R(6)  & 30 (2)   & 3.9 (1.2)\\
R(5)  & 41 (3)   & 4.7 (1.2)\\
R(4)  & 71 (4)   & 5.6 (1.0)\\
R(3)  & 71 (4)   & 5.2 (0.8)\\
R(2)  & \ldots   & 13.7 (1) \\
R(0)  & 55 (3)   & 10 (2)   \\
P(1)  & 55 (2)   & 6.4 (1.5)\\
P(2)  & 56 (3)   & \ldots      \\
P(3)  & 74 (2)   & \ldots      \\
P(4)  & 27 (2)   & \ldots      \\
P(18) & $<$15    & \ldots      \\
P(19) & 3.4 (0.4)& \ldots      \\
P(21) & 2.5 (0.8)& \ldots      \\
P(22) & $<6$     & \ldots      \\
\noalign{\smallskip} 
\tableline
\multicolumn{3}{p{4cm}}{ $^{\rm a}$ transitions not listed have not
been observed, are in a poor atmospheric region, or are blended with a
brighter isotope}\\
\multicolumn{3}{p{4cm}}{ $^{\rm b}$ $1\sigma$ errors in brackets}\\
\end{tabular}
\end{table}

\begin{figure}
\includegraphics[angle=90, scale=0.47]{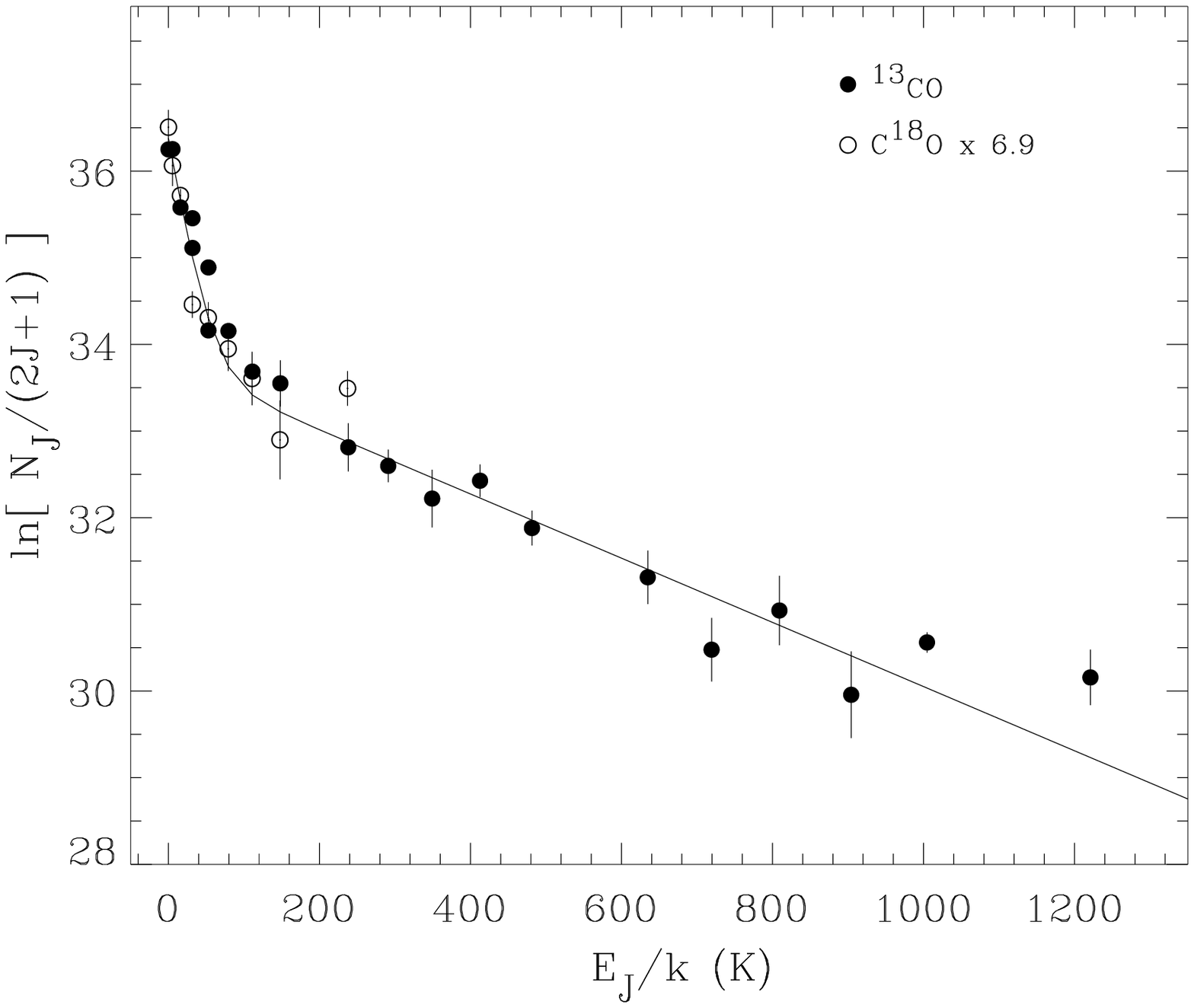}
\caption{Rotation diagram of \thirteenco\ (filled circles) and
\eighteenco\ (open circles) absorption lines toward \irsnine.  The
\eighteenco\ column densities have been scaled up by a factor of 6.9
to match the abundance of interstellar \thirteenco\ \citep{wil94}.
The good agreement of these isotopes in this plot implies that the
\thirteenco\ absorption lines are optically thin. The solid line
represents a two component gas with rotational temperatures of 20 and
270 K both at a \thirteenco\ column density of 4.65 $\times 10^{16}$
\sqcm.  Note that the 20 K gas absorbs at the systemic velocity, while
the warm component absorbs at much higher velocity
(Fig.~\ref{f:velo}).}~\label{f:rotco}
\end{figure}

At first glance, the line profiles of the fundamental ro-vibrational
CO transitions at 4.7 \mum, as published in \citet{mit90} and
\citet{boo02}, look similar to those of CH$_4$.  They show P Cygni
profiles as well, in \twelveco, \thirteenco, and perhaps even the
\eighteenco\ isotope, with the strength of emission versus absorption
increasing for higher $J$ levels. Upon closer inspection, however, the
absorption profiles of CO and CH$_4$ differ.  The low-$J$ isotopic CO
lines peak at 0 \kms, compared to --6 \kms\ for CH$_4$
(Fig.~\ref{f:velo}a). Higher-$J$ CO lines ($E_{\rm low}\geq 100$ K),
on the other hand, are dominated by absorption at blue-shifted
velocities $\geq 16$ \kms, which is absent in CH$_4$
(Fig.~\ref{f:velo}b). This provides clues to the formation and
evolution history of CH$_4$ (\S~\ref{sec:disc1}), and in this context
it is relevant to quantify the physical conditions of the absorbing
and emitting CO gas as well (Table~\ref{t:col}).

We analyzed the R=25,000 Keck/NIRSPEC spectrum of \citet{boo02} and
arrive at different conclusions compared to Mitchell et al. (1990,
1991).  First, and most importantly, in the Keck/NIRSPEC data
blue-shifted absorption at 16 \kms\ is seen up to $J$-levels as high
as 20, but only up to $J=3$ in \citet{mit91}.  In the rotation diagram
constructed from the integrated \thirteenco\ absorption equivalent
widths, this high velocity gas is responsible for the linear part
above energies of 100 K (Fig.~\ref{f:rotco}; Table~\ref{t:co}).  From
this a column density of $N$(\twelveco)=3.5$\pm 1.0 \times 10^{18}$
\sqcm\ at a temperature of $T_{\rm rot}=270\pm 100 $ K (3$\sigma$) is
derived. The column density is comparable to that found by
\citet{mit91}, but $T_{\rm rot}$ is a factor of 10 higher in the
Keck/NIRSPEC data. A second difference is that \eighteenco\ lines are
detected in the high S/N Keck/NIRSPEC data. The
\thirteenco/\eighteenco\ absorption line ratios of corresponding $J$
levels are equal to the interstellar isotope ratio of 6.9 at large
Galactocentric radii \citep{wil94} within a 20\% error margin (see the
\eighteenco\ data points in the \thirteenco\ rotation diagram;
Fig.~\ref{f:rotco}). An important consequence is that reliable CO
column densities can be derived, and thus reliable CH$_4$ abundances
in the various velocity ranges (\S4.1).  The steep part of the
\thirteenco\ rotation diagram, corresponding to gas at velocities of
--8 to +11 \kms, indicates $T_{\rm rot}=$20$\pm 7$ K and a column
density of $N$(\twelveco)=3.5$\pm 1.0 \times 10^{18}$ \sqcm, which is
a factor of 4 less compared to that found by \citet{mit90}. This is
due to our factor of 2 smaller measured equivalent widths and,
presumably, lower assumed line opacity.

For the emission lines, the observed \twelveco/\thirteenco\ peak
strength ratio is $6\pm 2$ at intermediate upper $J$ levels of 7-15.
Using the standard radiative transfer equation, the peak optical depth
of the \twelveco\ emission lines, $\tau_{12}$ can be derived from

\begin{equation}
\frac{F(^{12}{\rm CO})}{F(^{13}{\rm CO})}=\frac{1-{\rm
e}^{-\tau_{12}}}{1-{\rm e}^{-\tau_{12}/80}}.~\label{eq:tau}
\end{equation}

\noindent At a \twelveco/\thirteenco\ isotope ratio of 80
\citep{boo02}, it follows that $\tau_{12}=13\pm 4$.  Such an optical
depth corresponds to a column density of $N$(\twelveco)=$3.2\pm 1.0
\times 10^{18}$ \sqcm\ at an intrinsic line width of \bdop=3 \kms.  We
further discuss the origin of the emitting gas in \S\ref{sec:disc2}.

\subsection{The Solid CH$_4$ Absorption Bands}~\label{sec:ice}

\begin{figure}
\includegraphics[angle=90, scale=0.42]{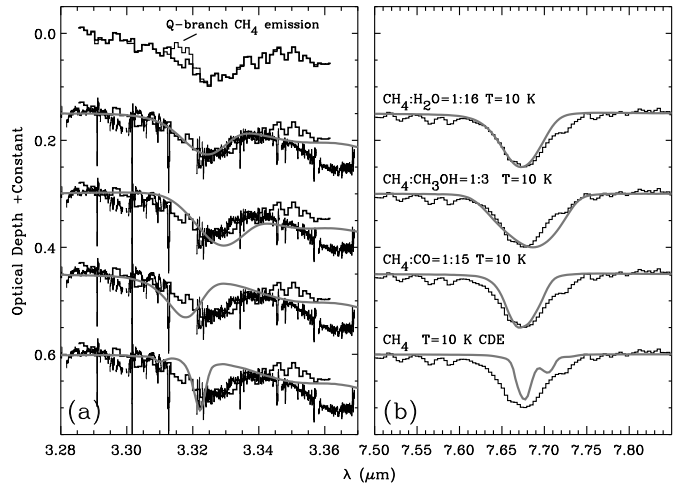}
\caption{The stretching mode of solid CH$_4$ observed with ISO/SWS
(thick histogram) and KECK/NIRSPEC (thin histogram), compared to a
number of laboratory ice simulations in panel (a). At the very top the
effect of removing the Q-branch emission of gas phase CH$_4$ is shown.
The CH$_4$ bending mode in panel (b) is fitted with the same
laboratory ice, without scaling the peak optical depth. Best fits are
obtained with H$_2$O--rich ices. Note the poor fit of CO--rich ices to
the CH$_4$ stretching mode.}~\label{f:ice}
\end{figure}

Solid CH$_4$ has been detected by its C-H bending mode at 7.67 \mum\
in ISO/SWS observations of \irsnine\ \citep{boo96}.  The peak position
and width of the interstellar band are well fitted with an amorphous,
low temperature (10 K) laboratory ice in which CH$_4$ is diluted in an
H$_2$O matrix. This same laboratory ice gives an excellent fit to the
new 3.32 \mum\ absorption feature detected in the Keck/NIRSPEC and
ISO/SWS spectra, not only in peak position and width, but also to the
peak optical depth of 0.1. We thus attribute this interstellar band in
its entirety to the C-H stretching mode of solid CH$_4$ in an
H$_2$O--rich ice (Fig.~\ref{f:ice}). A mismatch with the observed
ISO/SWS spectrum on the low wavelength wing is attributed to emission
by the strong Q-branch of gaseous CH$_4$ (Fig.~\ref{f:ice}). Indeed, a
better match is obtained after subtracting the model constructed from
the observed P and R-branch lines from the ISO/SWS spectrum (\S
3.2.1). In contrast, contamination of gas phase emission to the 7.67
\mum\ ice band is small because the continuum emission at this
wavelength is more than an order of magnitude brighter compared to the
continuum at 3.32 \mum\ \citep{whi96}.  Indeed, a weak inflection at
7.66 \mum\ is due to Q-branch {\it absorption}. Simultaneous fits of
the C-H stretch and bend modes highlight the good fit of CH$_4$:H$_2$O
ices (Fig.~\ref{f:ice}).  Pure CH$_4$ ices and mixtures with CO,
otherwise highly abundant in this line of sight \citep{tie91, boo02},
can be excluded at a high confidence level.  Note that the C-H
stretching mode of CH$_4$ better discriminates between polar and
apolar ices than does the bending mode.  Mixtures with other
high-dipole molecules, such as NH$_3$ and CH$_3$OH, do not provide
equally good fits compared to H$_2$O, but at present we cannot exclude
combinations of these polar species.

\section{Discussion}~\label{sec:disc}

\begin{table*}
\center
\caption{CH$_4$ and CO Physical Parameters toward \irsnine}~\label{t:col}
\begin{tabular}{lllccccl}
\noalign{\smallskip} 
\noalign{\smallskip} 
\tableline
\tableline
\noalign{\smallskip} 
Species        & Location$^{\rm a}$  & Phase$^{\rm b}$& $V^{\rm c}$&$T_{\rm ex}^{\rm d}$& $N^{\rm d}$ & $N/N_{\rm H}$$^{\rm d,e}$ & Notes  \\
               &           &          & \kms       &     K      & $10^{17}$ \sqcm\    &  10$^{-6}$    &                     \\
\noalign{\smallskip}  
\tableline
\noalign{\smallskip} 
CH$_4$         & outflow   & gas/abs. & $-$16      & 270        & $<0.1$              &  $<0.4$       & assumed $T$                        \\
CH$_4$         & core      & gas/abs. & $-$6       & 55$\pm$15  & 0.3$\pm$0.1         &  $\geq 3$     & \bdop=0.5--3 \kms\                 \\
CH$_4$         & envelope  & gas/abs. & 0          & 20         & $<0.1$              &  $<0.3$       & assumed $T$                        \\
CH$_4$         & core      & gas/em.  & +8         & 600$\pm$100& 0.58$^{+1.6}_{-0.3}$&  1.9$^{+7.2}_{-0.9}$& $4.3\times 10^{-14}$ ster$^{\rm f}$ \\
CH$_4^{\rm g}$ & envelope  & ice/abs. & \ldots     & \ldots     & 1.3$\pm 0.1$        &  2.1$\pm 0.7^{\rm h}$& \ldots                 \\
CO             & outflow   & gas/abs. & $-$16      & 270$\pm$100& 35$\pm$10           &  100          & opt. thin \thirteenco              \\
CO             & core      & gas/abs. & $-$6       & 55         &  $<10$              &  100          & assumed $T$                        \\
CO             & envelope  & gas/abs. & 0          & 20$\pm$7   &  35$\pm$10          &  100          & opt. thin \thirteenco              \\
CO             & core      & gas/em.  & +8         & 500$\pm$150&  30$\pm$10$^{\rm i}$&  100          & $4.3\times 10^{-14}$ ster$^{\rm f}$ \\
CO$^{\rm j}$   & envelope  & ice/abs. & \ldots     & \ldots     & 18$\pm$2            &   36$\pm 10^{\rm h}$& \ldots                             \\
\noalign{\smallskip} 
\tableline
\multicolumn{8}{p{12cm}}{ $^{\rm a}$ assumed location from velocity,
see also \S\S\ref{sec:disc1} and \ref{sec:disc2}; $^{\rm b}$ in gas or
ices, in emission or absorption; $^{\rm c}$ velocity w.r.t. systemic
$V_{\rm lsr}$=57 \kms; $^{\rm d}$ errors are 3$\sigma$; $^{\rm e}$
assuming $N_{\rm H}$/CO=1$\times 10^4$ from \citealt{lac94}; $^{\rm
f}$ solid angle derived from CO emission; $^{\rm g}$ \citealt{boo98};
$^{\rm h}$ ice abundance with respect to CO gas+ice abundance in
envelope; $^{\rm i}$ uncertainty in column dominated by uncertainty in
line opacity; $^{\rm j}$ \citealt{boo02}}
\end{tabular}
\end{table*}

\subsection{Astrochemical Implications}~\label{sec:disc1}

The detections of both the 3.32 \mum\ stretch and the 7.67 \mum\ bend
modes firmly establish the presence of both gaseous and solid CH$_4$
in the interstellar medium.  The gas+solid state abundance averaged
along the line of sight of the massive protostar \irsnine\ is
X[CH$_4$]=$2\times 10^{-6}$, and the absorption gas/solid state
abundance ratio is 0.23.  A combined high resolution CO/CH$_4$ line
profile analysis shows that the abundance as well as the gas/solid
state abundance ratio of CH$_4$ vary {\it locally} by an order of
magnitude, however (Table~\ref{t:col}). An ice+gas abundance of a few
$10^{-6}$ is found in the outer envelope (systemic velocity), the
inner envelope (--6 \kms) and the compact emission component (+8
\kms).  In the outer envelope, at 20 K the coldest component along the
line of sight, at least 90\% of the CH$_4$ is frozen on grains.
Within the envelope a segregation is present as well; CH$_4$ is
intimately mixed with H$_2$O and CH$_3$OH rather than with CO.  In the
warm, low velocity inner regions of \irsnine, CH$_4$ is mostly present
in the gas phase.  These enhanced abundances strongly contrast to the
absence of CH$_4$ in the massive, high velocity --16 \kms, warm
($T\simeq270$ K) outflow seen in CO (X[CH$_4$]$<0.4\times 10^{-6}$).

These new results provide clues to the formation and evolutionary
history of CH$_4$, refining previous scenarios \citep{lac91,
boo98}. In time dependent pure gas phase models, such as the
`standard' dense cloud model of \citet{lee96} and the UMIST model of
\citet{mil97}, gas phase CH$_4$ abundances of a few 10$^{-6}$ are
achieved after 10$^5$ years. Although this is comparable to the
observed abundances, there are strong arguments against a gas phase
origin: the low gas/solid state ratio and the intimate mixture between
CH$_4$ and H$_2$O in the ices in the envelope.  For comparison, CO, a
typical gas phase product, has an order of magnitude larger gas/solid
phase ratio compared to CH$_4$, even in the cold gas phase. H$_2$O on
the other hand is formed by hydrogenation of oxygen on the grains at
low temperatures (e.g. \citealt{jon84}), and a low gas/solid state
ratio is observed \citep{boo03}. It is thus likely that CH$_4$ was
formed on grain surfaces by hydrogenation of atomic C, at the same
time as H$_2$O was formed from atomic O. The low observed CH$_4$
abundance points to a formation time when most of the C was already
locked up in CO \citep{boo98}.

The gas phase CH$_4$ abundance in the --6 and +8 \kms\ components is
comparable to the solid state abundance in the envelope. The CH$_4$
may thus have been released from the grains in the inner regions near
the protostar. Because CH$_4$ is diluted in an H$_2$O--rich ice
(Fig.~\ref{f:ice}), its sublimation temperature is close to that of
H$_2$O, 90 K \citep{tie96, hir98}. The temperature of the +8 \kms\
emission component is indeed well above 90 K.  The gas at --6 \kms,
however, has a temperature of 55$\pm$15 K only.  While in one massive
protostar, W 33A, CH$_4$ gas at $T=110$ K was detected \citep{boo98},
a lower temperature of 50 K was also measured in another line of
sight, GL 7009S \citep{dar98}. We speculate that the icy mantles did
not thermally evaporate as a result of protostellar radiative heating,
but by a mild shock that has briefly heated the grains above 90
K. Subsequently the gas has cooled radiatively or adiabatically as the
inner \irsnine\ core expands outwards with a (current) velocity of 6
\kms.

Shocks may have had a profound impact on the higher velocity
gas. CH$_4$ is absent at velocities of 16 \kms\ and above, where warm
CO ($T=280$ K) is very abundant.  After release from the grains,
CH$_4$ must have efficiently chemically reacted. The observed CO
velocities up to 35 \kms\ point to the presence of non-destructive
C-type shocks.  These shocks have been well studied in the context of
molecular chemistry. At initial densities of 10$^5$ \cubcm, shocks
with velocities of 15--35 \kms\ have peak temperatures of 500-2500 K
(Bergin, Neufeld, \& Melnick 1998). At such high temperatures, and at
the low expected shock pressures\footnote{$log_{10}(P)=-13$ bar
assuming a factor of 2 density enhancement at a maximum temperature of
2500 K for a 35 \kms\ shock \citep{kau96, ber98}}, CH$_4$ is rapidly
driven into CO \citep{pri89}. The occurrence of a C shock therefore
explains the absence of CH$_4$ in the outflow of \irsnine.  The
importance of shock chemistry in high mass cores was deduced from low
gas/solid CO$_2$ abundance ratios as well \citep{cha00}, and by the
abundance ratios of S bearing species \citep{hat02}. One distinct
characteristic of shock chemistry is the formation of H$_2$O at
temperatures above 250 K (e.g. \citealt{ber98}). H$_2$O columns
comparable to CO are expected, i.e.  2.8$\times 10^{18}$ \sqcm\ for
the \irsnine\ outflow (Table~\ref{t:col}). This is a factor of 3.5
larger than the upper limit to the gas phase H$_2$O abundance derived
from ISO/SWS 6 \mum\ spectra \citep{boo03}.  High resolution 6 \mum\
spectra are required to further investigate the presence of H$_2$O in
the line of sight of \irsnine, the effect of blended emission and
absorption components and the relation to shocks and outflows.

\subsection{Nature of the Low Velocity Warm Gas}~\label{sec:disc2}

The elevated temperature of the low velocity, 6-8 \kms, absorption and
emission components seen both in CO and CH$_4$ locates this gas in the
inner regions of the protostellar core. The present observations
further constrain the origin of this material.

The area of the radiating physical body can be calculated by matching
the \twelveco\ column density derived from the isotope ratio
(\S~\ref{sec:co}) to the emission line fluxes. The measured \twelveco\
line fluxes need to be corrected for continuum absorption by
foreground material. We derive a 4.7 \mum\ continuum extinction
magnitude of $A_{\rm 4.7}=1.8\pm 0.6$ using the interstellar
extinction law of \citet{dra03} and an optical extinction $A_{\rm V}$
of 60$\pm$20. The latter is derived from the $A_{\rm V}$/$N_{\rm H}$
relation of \citet{boh78} and the average line of sight hydrogen
column density from this work (Table~\ref{t:col}) and that from the
silicate absorption band \citep{tie91}. The extinction corrected line
fluxes are thus matched at a solid angle of 4.3$\times 10^{-14}$
sterad. At the distance to \irsnine\ of 2.8 kpc \citep{cra78} this
corresponds to a radius of 70 AU. The similarity of CO and CH$_4$
velocities and emission line widths (Fig.~\ref{f:velo}c) indicates
that both molecules are present in this same volume.

The CO and CH$_4$ gas detected on these scales may well be related to
the dense material that was inferred on scales of $\leq 300-600$ AU
from unresolved 3 millimeter continuum interferometer images
\citep{tak00}.  This intense emission is unrelated to the envelope,
and instead may originate from a dense shell or circumstellar disk.
The P Cygni-like line profiles in both CO and CH$_4$ may suggest that
an expanding envelope is the preferred geometry. In these line
profiles the absorption component is red-shifted by the shell
expansion velocity, and the emission component is centered at the
systemic velocity and is broadened by twice the expansion
velocity. \citet{mit91} argue that the CO emission component is
blue-shifted due to the effect of deep absorption by foreground
material at the systemic velocity. This hypothesis can be tested by
the profiles of the highest $J$ transitions, which should not have
foreground contributions. We find that even the lines from the highest
rotational levels of CO are blue shifted by +8 \kms\
(Fig.~\ref{f:velo}c). Furthermore, the CH$_4$ emission lines peak at
this same velocity and are also unresolved. Perhaps the \irsnine\ core
moves within its envelope by 8 \kms\ and the observed peak of the
emission line is the center of the P Cygni profile.  In this scenario
the shell expansion velocity is 14 \kms\ and an emission line width of
28 \kms\ is expected.  Such broad lines are inconsistent with the
observations as well.  We conclude that the line profiles are not true
P Cygni type profiles and thus are inconsistent with a spherically
symmetric expanding envelope.  It is unclear what the relation is
between the 55 K absorbing gas seen at $-$6 \kms\ and the 500 K
emitting gas at +8 \kms.  More detailed dynamical models of
circumstellar disks, accretion flows and outflows are required. The
possibility of photon excitation needs to be considered as well.  As
noted in \citet{mit91} the similarity of CO absorption and emission
line profiles between a number of massive protostars suggests that
these lines trace a fundamental property of their inner cores. This is
strengthened by the similarity of the CH$_4$ line profiles.

\section{Conclusions and Future Work}~\label{sec:concl}

Using high resolution L and M band spectroscopy we have shown that the
gas phase CH$_4$ abundance varies considerably in the line of sight of
the massive protostar \irsnine. This is a result of specific molecule
formation and destruction processes: CH$_4$ is formed on grain
surfaces and subsequently evaporated from the grains in the inner
regions of the protostar.  Both evaporation by photon heating and by
shocks likely have played roles.  CH$_4$ efficiently `burns' to CO in
the highest velocity shocked regions. The gas phase CH$_4$/CO
abundance ratio is thus a sensitive tracer of shock chemistry. High
spectral resolution observations of other species, in particular
H$_2$O at 6 \mum, are required to further test the importance and
nature of shock chemistry in high mass star forming regions. The {\it
Stratospheric Observatory For Infrared Astronomy (SOFIA)} will offer
this opportunity in the near future \citep{lac02}. Observations of the
3.32 \mum\ CH$_4$ stretch mode in more lines of sight require high
radial velocities ($>30$ \kms) or temperatures that populate the
highest $J$ levels to avoid saturated telluric lines. In this respect,
the 7.67 \mum\ bending mode may be more suited.  Future observations
of CH$_4$ toward low mass protostars are particularly exciting, as the
CH$_4$/CO ratio is a sensitive tracer of protoplanets. In the
relatively low pressure solar nebula low CH$_4$/CO abundances
occur. The higher pressure in protoplanetary sub-nebulae, however,
favors higher CH$_4$/CO ratios \citep{pri89}. Variations of the CH$_4$
abundance between solar system comets possibly trace such physical
condition gradients in the early protoplanetary disk \citep{gib03}.

\acknowledgments

The research of A.C.A.B. and G.A.B. is supported by the Spitzer Legacy
Science program and by the Owens Valley Radio Observatory through NSF
grant AST-0228955. The authors wish to recognize and acknowledge the
very significant cultural role and reverence that the summit of Mauna
Kea has always had within the indigenous Hawaiian community.  We are
most fortunate to have the opportunity to conduct observations from
this mountain. We thank Jan Cami for providing a copy of ISO/SWS IA,
Fred Lahuis for an independent reduction, Tim Brooke for suggesting
the possible ethane identification, and Yuk Yung for help with the
wavelength calibration.  Suggestions by an anonymous referee have led
to an improved presentation of the data.


\begin{thebibliography}{}
\bibitem[Alexander et al.(2003)]{ale03} Alexander, R.~D., Casali,
 M.~M., Andr{\' e}, P., Persi, P., \& Eiroa, C.\ 2003, \aap, 401, 613
\bibitem[Allamandola et al. (1992)]{all92} Allamandola, L.J, Sandford,
 S.A., Tielens, A.G.G.M., \& Herbst, T.M. 1992, \apj, 399, 134
\bibitem[Bergin et al.(1998)]{ber98} Bergin, E.~A., Neufeld, D.~A., \&
 Melnick, G.~J.\ 1998, \apj, 499, 777
\bibitem[Bohlin, Savage, \& Drake(1978)]{boh78} Bohlin, R.~C., Savage,
 B.~D., \& Drake, J.~F.\ 1978, \apj, 224, 132
\bibitem[Boogert et al.(1996)]{boo96} Boogert, A.~C.~A., et al.\ 1996,
 \aap, 315, L377
\bibitem[Boogert et al.(1997)]{boo97} Boogert, A.~C.~A., Schutte,
 W.~A., Helmich, F.~P., Tielens, A.~G.~G.~M., \& Wooden, D.~H.\ 1997,
 \aap, 317, 929
\bibitem[Boogert et al.(1998)]{boo98} Boogert, A.~C.~A., Helmich,
 F.~P., van Dishoeck, E.~F., Schutte, W.~A., Tielens, A.~G.~G.~M., \&
 Whittet, D.~C.~B.\ 1998, \aap, 336, 352
\bibitem[Boogert et al. (2000)]{boo00} Boogert, A.C.A., Ehrenfreund
 P., Gerakines, P.A., Tielens, A.G.G.M., Whittet, D.C.B., et al. 2000,
 \aap, 353, 349
\bibitem[Boogert et al.(2002)]{boo02} Boogert, A.~C.~A., Blake, G.~A.,
 \& Tielens, A.~G.~G.~M.\ 2002, \apj, 577, 271
\bibitem[Boogert et al.(2004)]{boo04} Boogert, A. C. A., et al.  2004,
 \apjs, 154, in press
\bibitem[Boonman \& van Dishoeck(2003)]{boo03} Boonman, A.~M.~S.~\&
 van Dishoeck, E.~F.\ 2003, \aap, 403, 1003
\bibitem[Boudin et al.(1998)]{bou98} Boudin, N., Schutte, W.~A., \&
 Greenberg, J.~M.\ 1998, \aap, 331, 749
\bibitem[Brooke, Sellgren, \& Smith(1996)]{bro96} Brooke, T.~Y.,
 Sellgren, K., \& Smith, R.~G.\ 1996, \apj, 459, 209
\bibitem[Brooke et al.(1999)]{bro99} Brooke, T.~Y., Sellgren, K., \&
 Geballe, T.~R.\ 1999, \apj, 517, 883
\bibitem[Brown, Charnley, \& Millar (1988)]{bro88} Brown, P.~D.,
 Charnley, S.~B., \& Millar, T.~J.\ 1988, \mnras, 231, 409
\bibitem[Brown et al.(2003)]{bro03} Brown, L.~R., et al.\ 2003,
 Journal of Quantitative Spectroscopy and Radiative Transfer, 82, 219
\bibitem[Ceccarelli et al.(2001)]{cec01} Ceccarelli, C., Loinard, L.,
 Castets, A., Tielens, A.~G.~G.~M., Caux, E., Lefloch, B., \& Vastel,
 C.\ 2001, \aap, 372, 998
\bibitem[Charnley \& Kaufman(2000)]{cha00} Charnley, S.~B.~\& Kaufman,
 M.~J.\ 2000, \apjl, 529, L111
\bibitem[Chiar et al.(2000)]{chi00} Chiar, J.~E., Tielens,
 A.~G.~G.~M., Whittet, D.~C.~B., Schutte, W.~A., Boogert, A.~C.~A.,
 Lutz, D., van Dishoeck, E.~F., \& Bernstein, M.~P.\ 2000, \apj, 537,
 749
\bibitem[Crampton, Georgelin, \& Georgelin (1978)]{cra78} Crampton,
 D., Georgelin, Y.~M., \& Georgelin, Y.~P.\ 1978, \aap, 66, 1
\bibitem[Dartois et al.(1998)]{dar98} Dartois, E., D'Hendecourt, L.,
 Boulanger, F., Jourdain de Muizon, M., Breitfellner, M., Puget, J.-L.,
 \& Habing, H.~J.\ 1998, \aap, 331, 651
\bibitem[Dartois et al.(1999)]{dar99} Dartois, E., Schutte, W.,
 Geballe, T.~R., Demyk, K., Ehrenfreund, P., \& D'Hendecourt, L.\ 1999,
 \aap, 342, L32
\bibitem[Dartois et al.(2002)]{dar02} Dartois, E., d'Hendecourt, L.,
 Thi, W., Pontoppidan, K.~M., \& van Dishoeck, E.~F.\ 2002, \aap, 394,
 1057
\bibitem[de Graauw et al.(1996)]{gra96} de Graauw, T.,~et al.\ 1996,
 \aap, 315, L49
\bibitem[Draine(2003)]{dra03} Draine, B.~T.\ 2003, \araa, 41, 241
\bibitem[Fischer et al.(2003)]{fis03} Fischer, J., Gamache, R.~R.,
 Goldman, A., Rothman, L.~S., \& Perrin, A.\ 2003, Journal of
 Quantitative Spectroscopy and Radiative Transfer, 82, 401
\bibitem[Genzel(1992)]{gen92} Genzel, R. 1992, in The Galactic
 Interstellar Medium, ed. D. Pfenniger \& P. Bartholdi (Berlin:
 Springer), 275
\bibitem[Gerakines et al. (1999)]{ger99} Gerakines, P.A., Whittet,
 D.C.B., Ehrenfreund, P., Boogert, A.C.A., Tielens, A.G.G.M., et
 al. 1999, \apj, 522, 357
\bibitem[Gibb et al.(2003)]{gib03} Gibb, E.~L., Mumma, M.~J., dello
 Russo, N., Disanti, M.~A., \& Magee-Sauer, K.\ 2003, Icarus, 165, 391
\bibitem[G{\" u}rtler et al.(2002)]{gue02} G{\" u}rtler, J., Klaas,
 U., Henning, T., {\' A}brah{\' a}m, P., Lemke, D., Schreyer, K., \&
 Lehmann, K.\ 2002, \aap, 390, 1075
\bibitem[Hatchell \& Viti(2002)]{hat02} Hatchell, J.~\& Viti, S.\
 2002, \aap, 381, L33
\bibitem[Hiraoka et al.(1998)]{hir98} Hiraoka, K., Miyagoshi, T.,
 Takayama, T., Yamamoto, K., \& Kihara, Y.\ 1998, \apj, 498, 710
\bibitem[Jones \& Williams(1984)]{jon84} Jones, A.~P.~\& Williams,
 D.~A.\ 1984, \mnras, 209, 955
\bibitem[Kaufman \& Neufeld(1996)]{kau96} Kaufman, M.~J.~\& Neufeld,
 D.~A.\ 1996, \apj, 456, 250
\bibitem[Kessler et al.(1996)]{kes96} Kessler, M.~F., et al.\ 1996,
 \aap, 315, L27
\bibitem[Knacke et al.(1985)]{kna85} Knacke, R.~F., Noll, K.~S.,
 Geballe, T.~R., \& Tokunaga, A.~T.\ 1985, \apjl, 298, L67
\bibitem[Lacy et al.(1991)]{lac91} Lacy, J.~H., Carr, J.~S., Evans,
 N.~J., Baas, F., Achtermann, J.~M., \& Arens, J.~F.\ 1991, \apj, 376,
 556
\bibitem[Lacy et al.(1994)]{lac94} Lacy, J.~H., Knacke, R., Geballe,
 T.~R., \& Tokunaga, A.~T.\ 1994, \apjl, 428, L69
\bibitem[Lacy et al.(2002)]{lac02} Lacy, J.~H., Richter, M.~J.,
 Greathouse, T.~K., Jaffe, D.~T., \& Zhu, Q.\ 2002, \pasp, 114, 153
\bibitem[Lee, Bettens, \& Herbst(1996)]{lee96} Lee, H.-H., Bettens,
 R.~P.~A., \& Herbst, E.\ 1996, \aaps, 119, 111
\bibitem[Lide(1994)]{lid94} Lide, D. R., 1994, CRC Handbook of
 Chemistry and Physics (75th ed.; Boca Raton: CRC Press)
\bibitem[Markwick, Millar, \& Charnley(2000)]{mar00} Markwick, A.~J.,
 Millar, T.~J., \& Charnley, S.~B.\ 2000, \apj, 535, 256
\bibitem[McLean et al. (1998)]{mcl98} McLean, I.S., Becklin, E.E.,
 Bendiksen, O., Brims, G., \& Canfield, J. 1998, Proc. SPIE, 3354, 566
\bibitem[Millar, Farquhar, \& Willacy(1997)]{mil97} Millar, T.~J.,
 Farquhar, P.~R.~A., \& Willacy, K.\ 1997, \aaps, 121, 139
\bibitem[Mitchell et al. (1990)]{mit90} Mitchell, G.F., Maillard,
 J.-P., Allen, M., Beer, R., \& Belcourt, K. 1990, \apj, 363, 554
\bibitem[Mitchell et al.(1991)]{mit91} Mitchell, G.~F., Maillard,
 J.-P., \& Hasegawa, T.~I.\ 1991, \apj, 371, 342
\bibitem[Noriega-Crespo et al.(2004)]{nor04} Noriega-Crespo, A., et al.
 2004, \apjs, 154, in press
\bibitem[Pendleton et al.(1994)]{pen94} Pendleton, Y.~J., Sandford,
 S.~A., Allamandola, L.~J., Tielens, A.~G.~G.~M., \& Sellgren, K.\
 1994, \apj, 437, 683
\bibitem[Pontoppidan et al.(2003)]{pon03} Pontoppidan, K.~M., Dartois,
 E., van Dishoeck, E.~F., Thi, W.-F., \& d'Hendecourt, L.\ 2003, \aap,
 404, L17
\bibitem[Prinn \& Fegley(1989)]{pri89} Prinn, R. G., \& Fegley,
 B. 1989, in: ``Origin and Evolution of Planetary and Satellite
 Atmospheres'', eds.  S. K. Atreya, J. B. Pollack, M. S. Matthews, The
 University of Arizona Press, p. 78
\bibitem[Rothman et al.(2003)]{rot03} Rothman, L.~S., et al.\ 2003,
 Journal of Quantitative Spectroscopy and Radiative Transfer, 82, 5
\bibitem[Sandford \& Allamandola (1988)]{san88} Sandford, S.A., \&
 Allamandola, L.J. 1988, Icarus, 76, 201
\bibitem[Sandford et al.(1991)]{san91} Sandford, S.~A., Allamandola,
 L.~J., Tielens, A.~G.~G.~M., Sellgren, K., Tapia, M., \& Pendleton,
 Y.\ 1991, \apj, 371, 607
\bibitem[Schutte et al.(1996)]{sch96} Schutte, W.~A., Gerakines,
 P.~A., Geballe, T.~R., van Dishoeck, E.~F., \& Greenberg, J.~M.\ 1996,
 \aap, 309, 633
\bibitem[Spitzer(1978)]{spi78} Spitzer, L. 1978, Physical Processes
 in the Interstellar Medium (New York: Wiley)
\bibitem[Spoon et al.(2001)]{spo01} Spoon, H.~W.~W., Keane, J.~V.,
 Tielens, A.~G.~G.~M., Lutz, D., \& Moorwood, A.~F.~M.\ 2001, \aap,
 365, L353
\bibitem[Tielens et al. (1991)]{tie91} Tielens, A.G.G.M., Tokunaga,
 A.T., Geballe, T.R., \& Baas, F. 1991, \apj, 381, 181
\bibitem[Tielens \& Whittet(1996)]{tie96} Tielens, A.~G.~G.~M.~\&
 Whittet, D.~C.~B.\ 1996, IAU Symp.~178: Molecules in Astrophysics:
 Probes \& Processes, 178, 45
\bibitem[van der Tak et al. (2000)]{tak00} van der Tak, F.F.S., van
 Dishoeck, E.F., Evans, N.J., \& Blake, G.A. 2000, \apj, 537, 283
\bibitem[White et al.(2000)]{whi00} White, G.~J., et al.\ 2000, \aap,
364, 741
\bibitem[Whittet et al. (1996)]{whi96} Whittet, D.C.B., Schutte, W.A.,
 Tielens, A.G.G.M., Boogert, A.C.A., de Graauw, T., et al. 1996, \aap,
 315, 357
\bibitem[Willner et al.(1982)]{wil82} Willner, S.~P., et al.\ 1982,
\apj, 253, 174
\bibitem[Wilson \& Rood (1994)]{wil94} Wilson, T.L., \& Rood,
 R.T. 1994, \araa, 32, 191
\end{thebibliography}
\end{document}